\documentclass{ws-procs975x65} 
\newcommand{\onlinecite}[1]{\hspace{-1 ex} \nocite{#1}\citenum{#1}}            
\begin{document}
\title{The ``glass transition'' as a topological defect driven transition in a distribution of crystals and a prediction of a universal viscosity collapse}

\author{Z. Nussinov$^*$ and N. B. Weingartner}
\address{Department of Physics and Institute of Materials Science and Engineering, Washington University, \\
St. Louis, Missouri 63130, USA \\
$^*$E-mail: zohar@wustl.edu,\\
weingartner.n.b@wustl.edu}

\author{F. S. Nogueira}
\address{Institute for Theoretical Solid State Physics,\\
 IFW Dresden, PF 270116, 01171 Dresden, Germany\\
E-mail: f.de.souza.nogueira@ifw-dresden.de}

\begin{abstract}
Topological defects are typically quantified relative to ordered backgrounds. The importance of these defects to the understanding of physical phenomena including diverse equilibrium melting transitions from low temperature ordered to higher temperatures disordered systems (and vice versa) can hardly be overstated.  Amorphous materials such as glasses seem to constitute a fundamental challenge to this paradigm. A long held dogma is that transitions into and out of an amorphous glassy state are distinctly different from typical equilibrium phase transitions and must call for radically different concepts. In this work, we critique this belief. We examine systems that may be viewed as simultaneous distribution of different ordinary equilibrium structures. In particular, we focus on the analogs of melting (or freezing) transitions in such distributed systems. The theory that we arrive at yields dynamical, structural, and thermodynamic behaviors of glasses and supercooled fluids that, for the properties tested thus far, are in qualitative and quantitative agreement with experiment. We arrive at a prediction for the viscosity and dielectric relaxations that is universally satisfied for all experimentally measured supercooled liquids and glasses over 15 decades. 
\end{abstract}

\keywords{Style file; \LaTeX; Proceedings; World Scientific Publishing.}

\bodymatter

\section{Conventional Melting via the Condensation of Topological Defects}

In their landmark studies, Berezinski  \cite{BKT1,BKT2} and Kosterlitz and Thouless  \cite{KT,KT'} investigated (BKT) transitions in two-dimensional (2D) systems \cite{Jose_book}. These systems (as nearly all other 2D theories) do not exhibit usual long range order at low temperatures, e.g., \cite{MW,Coleman,MS,zn_thesis}.  The vitally important concept of ``topological order'' that Kosterlitz and Thouless first introduced and coined \cite{KT} to aid the description of the transitions in the classical systems that they investigated has evolved over the years and is currently of use also for myriad singular properties of quantum systems at both zero \cite{wen_book} and finite temperatures \cite{NO}. Haldane \cite{Haldane}, Thouless and his collaborators \cite{Thouless}, and other pioneers discovered the novel role of topology in many other (often deeply interrelated) physical arenas including, notably, various quantum systems. It is a pleasure to write in this volume honoring the watershed contributions of Haldane, Kosterlitz, and Thouless that have been recognized by the 2016 Nobel Prize in Physics. The works of these three laureates ushered very rich applications of topology in condensed matter physics. Apart from providing conceptual breakthroughs in more exotic systems, topology has proven to be very instrumental in understanding everyday behaviors such as usual phase transitions. 

Typical melting transitions from solids to liquids proceed by the condensation of topological defects in a crystal (e.g., the appearance of grain boundaries, dislocations and disinclinations in ever increasing numbers and volume). Beyond a certain threshold, crystalline order yields to these defects. As they progressively proliferate, the defects ultimately eradicate any trace of crystalline order, rigidity is lost, and the system transitions into a fluid. This process is a simple extension of the condensation of vortices in the 2D models examined by Kosterlitz and Thouless. In expansive books \cite{Kleinert1,Kleinert2}, Kleinert fleshed out these notions in a detailed way. Kleinert cast the melting of crystals in a field theoretic framework that underscored how melting precisely occurs via the generation of topological defects. This approach has been extended to the quantum arena \cite{Beekman}. In numerous systems (including those studied by Kosterlitz and Thouless), the topological defects that restore the symmetry of the continuum fluid are, in a precise mathematical sense, ``dual'' to the fields that characterize standard orders. Conventional measures of topological defects often rely on a comparison between a physical configuration vis a vis an idealized one. For instance, dislocations in a crystal, are defined and quantified by their topological charge- the so called ``Burgers vector''  \cite{Kleinert1,nabarro}.  The Burgers vector is the total sum of the displacements of all atoms that lie on a contour surrounding a given lattice point. The atomic displacements that are measured here are those of the atoms in the deformed solid by
comparison to the locations of the very same atoms in a pristine ``ideal'' crystal that experiences no deformations. In a similar vein, vortices of pertinence to the
BKT transitions are defined relative to a uniform background.   

While these ideas are extremely alluring, there is, of course, far more to life than such idealized crystalline or uniform backgrounds. Commonplace amorphous low temperature materials such as glasses do not veer towards an idealized regular periodic crystalline array of atoms relative to which displacements may be measured in a meaningful way. Thus, the usual notion of defects seems to be somewhat ill-defined if one were to try to blindly apply it to such materials. This lack of clearly discernible defects is not merely an academic issue. The existence or absence of topological excitations carries very real practical consequences. As is well appreciated, a sufficiently stressed crystalline solid can falter and exhibit ``slip'' via the ``glide'' motion of dislocations (the movement of dislocations parallel to their Burgers vector). When an external force is applied to a crystal, dislocations may drift far more readily than the surrounding crystal. The ease with which defects such as dislocations may be made to move as opposed to pushing an entire crystal move {\em en masse} has a vivid analogy attributed to Orowan \cite{nabarro}
(a practical trick that according to legend inspired him to think of glide). This analogy is based on the fact that one may trivially move an extremely heavy carpet (a runner) by repeatedly creating a ruck- a small indentation- in the rug (a ``defect'') and then merely stepping on this indentation one foot at a time so as to push the ruck from one side of the carpet to the other. Repeatedly marching, rather effortlessly, on the carpet in this way leads to a net displacement of the carpet. In a somewhat similar manner, point like defects may be pushed by an external force and ultimately lead to motions of entire atomic planes. Finding defects may thus enable the delineation of unstable regions in the solid when an external force is applied. There are field theoretic \cite{EPJE} as well as unsupervised \cite{EPJE,Scientific_reports} and supervised \cite{softness} machine learning type approaches to define and ascertain defects and structures in amorphous media. Additional measures in amorphous media include the determination of length scales associated with the penetration of external shear \cite{nick1,nick2}. However, generally, in glasses, where defects are not recognizable in the usual way, there are no evident ``weak spots'' that may, so readily, respond to external forces as they do in crystals.

There are numerous metallurgical advantages associated with the lack of usual topological defects in amorphous materials. Indeed, thanks to the absence of sharp topological defects, metallic glasses exhibit extreme strength and hardness, are very fracture resistance, and are exceedingly elastic \cite{Telford}. These practical benefits, however, come with a prohibitive price tag for theories that aim to explain the transition out of or into the glassy state via conventional melting or freezing. Indeed, a celebrated drawback of the dearth of sharply defined topological defects is the inability to rationalize the transition from rapidly supercooled liquids into the complex (and seemingly largely random) structure of the glass and vice versa by the proliferation of defects that destroy usual order. 

\section{Freezing or Melting of Simultaneous Equilibrium Phases} 

\begin{figure}[h]
\begin{center}
\includegraphics[width=5in]{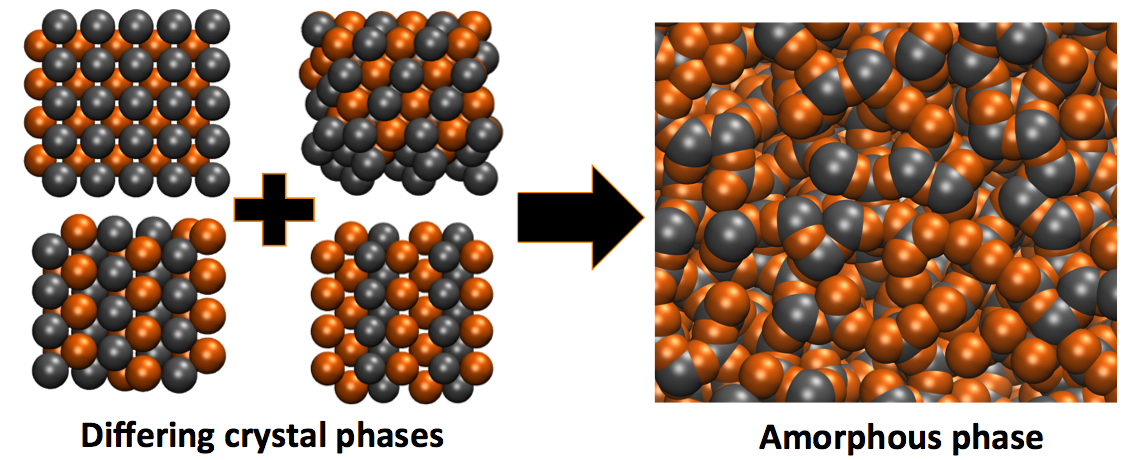}
\end{center}
\caption{From \cite{Perim}.
A multitude of equilibrium crystal phases of similar energies (left)
may ``entangle''.
This thwarts crystallization and leads to complex amorphous structures. 
process.}
\label{fig:perim}
\end{figure}

In this article, we will consider systems of $N \gg 1$ atoms occupying a volume $V$ of size ${\cal{O}}(N)$. The energy density $\epsilon \equiv E/V$ with $E$ the total system energy. The central question that we focus on is whether glasses may be investigated by a rather trivial extension of the above ideas concerning melting via the proliferation of topological defects. Specifically, we wish to ask what will transpire for an ensemble average of different equilibrium systems each of which may, on its own, ``freeze'' into an ordered crystal at low temperatures (or energy densities). An illustration of this idea is provided in Fig. \ref{fig:perim} (adopted from \cite{Perim}- an article further building on the notion of ``confusion'' introduced in \cite{Confusion}). Our concept differs from that of ``confusion'' \cite{Confusion}. However, the resulting structural picture that our concept gives rise to is similar. The basic premise is that there are multiple equilibrium states of similar energy density. When the system is supercooled, these states of nearly identical energies may appear in unison. A ``superposition'' of such equilibrium states general can lead to the observed amorphous structures. In 
Sections \ref{sec:collapse} and thereafter we will analyze what will occur in the combined system if each of the individual states exhibits its own freezing/melting transition. This ``hodge-podge" of overlapping states contributes to an internal frustration which greatly influences the dynamics and thermodynamics of the system, chief among them the dynamic viscosity. This analysis leads to a new empirically observed collapse for the viscosities of all glass formers.

\section{A Lightning Review of Transitions into the Glassy State} 
\label{sec:glass}

\begin{figure}
\centering
\includegraphics[width=6.2in]{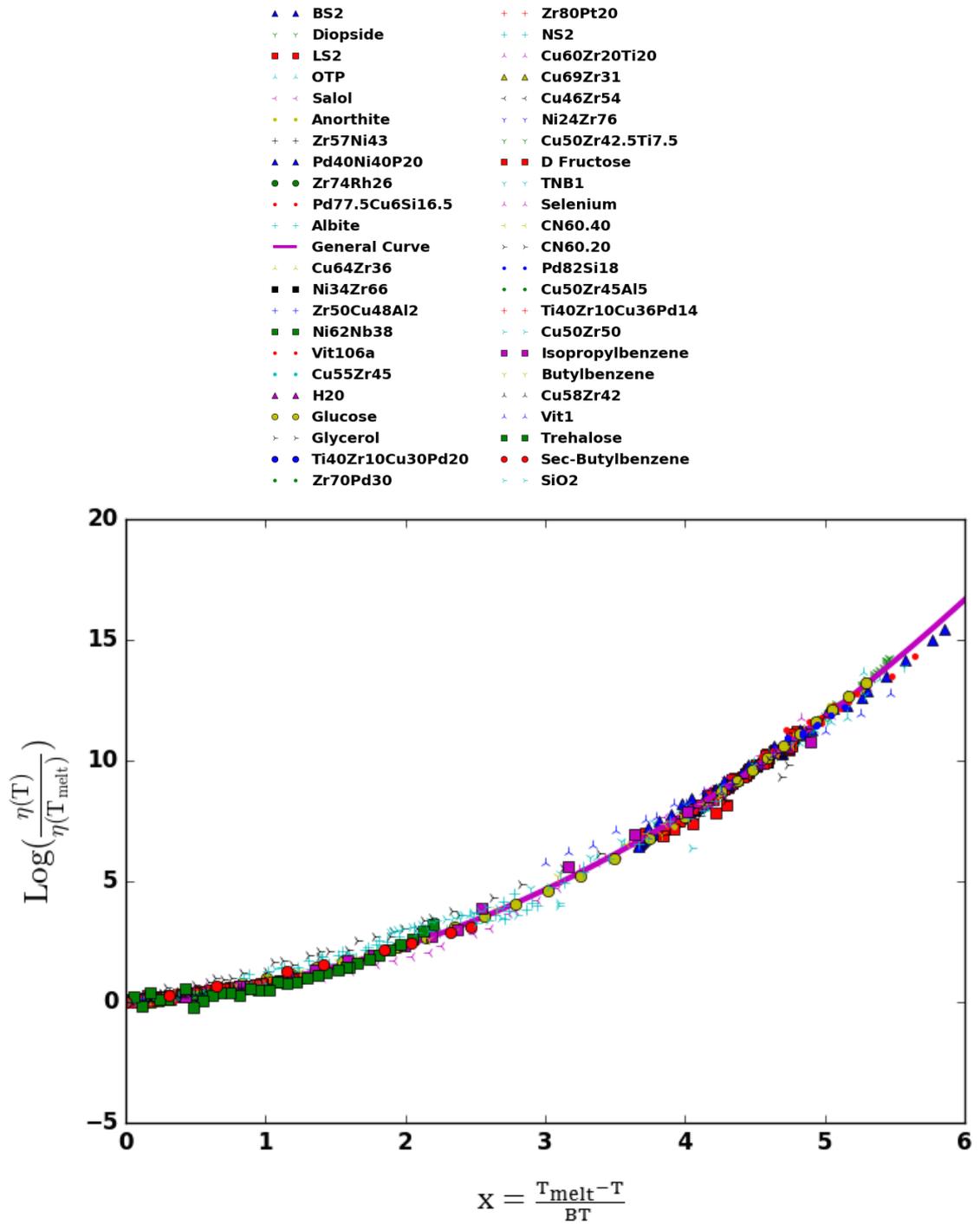}
\caption{\small \it 
{\em The scaling relation of Eq. (\ref{scaling}).  The``General Curve'' is given by the prediction of Eq. (\ref{fz}).}}
\label{fig:collapse}
\end{figure}

Before proceeding further, we very briefly regress to a discussion concerning the ubiquitous nature of glasses and their striking nature. Glasses are one of the most common states of matter  \cite{Telford,bib:tech1,bib:tech2,bib:tech3,bib:tech4,bib:tech6,bib:paw,bib:rmp,bib:review}. Nonetheless, even after millennia of use, they still remain poorly understood. In principle, {\it any} system may be made glassy by supercooling. ``Supercooling'' refers to a rapid cooling of a liquid below its melting (or freezing) temperature so that crystallization has insufficient time to occur. It is important to emphasize that these systems are {\em disorder free}. As the temperature of supercooled liquids is further  lowered below melting, they become increasingly sluggish. At low enough temperatures, when the viscosity ($\eta$) biomes larger than the (rather arbitrary) threshold value of $10^{12}$ Pascal $\times$ second, the resulting system is christened a ``glass''. At temperatures lower than the ``glass transition temperature'' $T_{g}$ at which this threshold is obtained, the relaxation times become too long to be easily measured. Glasses exhibit solid-like rigidity on measurable time scales. However, by comparison to ordered equilibrium solids, glasses are amorphous. The ``transition'' of liquids into the glassy state is unusual. Typical systems display significant  changes in all measurable quantities when they transition from one phase to another. This, however, is not the case for glasses. While the dynamics of supercooled liquids typically display significant changes (the viscosity and relaxation times of numerous supercooled liquids increase by many orders of magnitude prior to becoming glasses), \cite{bib:rmp,bib:review,bib:angell} static properties (such as structure and thermodynamic observables) may exhibit very modest changes.

The most prevalent (Vogel-Fulcher-Tamann-Hesse (VFTH)) \cite{bib:vft} fit for the viscosities of supercooled liquids is, by now, almost a century old. This fit states that, as a function of the temperature $T$, the viscosity $\eta$ of a supercooled liquid is given by $\eta(T) = \eta_{0} e^{DT_{0}/(T-T_{0})}$. Here, $T_{0}, D,$ and $\eta_{0}$ are system specific constants. Taken literally, this fit implies a singular temperature ($T_{0}$) at which the viscosity diverges. In tandem with this physical interpretation, $T_{0}$ is often called the ``ideal glass transition temperature''. Since there are very few measurements of the viscosity below $T_{g}$, it has been hard to critically assess this putative divergence. All known fits and associated theories of supercooled liquids and glasses argue for various temperature (and other) dependences unrelated to the simple equilibrium melting transition temperature, e.g., \cite{bib:rmp,bib:review,bib:ag1,bib:ag2,bib:gt5,bib:CG,bib:mct1,bib:mct2,bib:mct3,bib:KKZNT1,bib:KKZNT2,bib:KKZNT3,bib:BENK1,bib:BENK2,bib:Blodgett,bib:MYEGA}. These often appear as ad hoc additions to existing frameworks. In essence, the existence of the glassy state has always been ascribed to particular processes that are radically different from those in standard {\it equilibrium melting (or freezing) transitions}. 

\section{A Universal Collapse of Experimental Viscosity Data Over 15 Decades}
\label{sec:collapse}

The theoretical approach to the glass transition, \cite{bib:DEH1,bib:DSH,bib:DEH2,QZN} based on the extension 
of defect melting to an ensemble of {\it equilibrium states}, that we will review in the upcoming sections motivated us to investigate the viable existence of a universal collapse of the viscosity and other data in which the standard {\it equilibrium melting transition temperature} $T_{melt}$ (rather specifically, the ``liquidus'' temperature above which the equilibrium system is its liquid state) is the dominant temperature. Thus, we ask whether the viscosities of supercooled liquids and glasses might be governed by the common equilibrium melting transition temperature. If the answer to this question is affirmative then there might not be a need to introduce singular temperatures (such as $T_{0}$), appeal to the possible character of activation processes in complex high dimensional energy landscapes, or invoke other assumptions in order to rationalize the increase of the viscosities of supercooled liquids. With this in mind, we studied \cite{bib:DSH} 
45 supercooled liquids of all known varieties (including silicate, organic, chalcogenide, and metallic glassformers) to discern if the dependence of the viscosity on temperature is governed by the equilibrium melting temperature. Specifically, we asked whether 
\begin{eqnarray}
\frac{\eta(T)}{\eta(T_{melt})} = F( \frac{T_{melt}-T}{B T}),
\label{scaling}
\end{eqnarray}
with $F(x)$ a universal function and $B$ is a material-dependent (dimensionless) constant. The results of our analysis are displayed in Fig. \ref{fig:collapse}. The observed collapse demonstrates that Eq. (\ref{scaling}) is satisfied. The values of $B$ (tabulated towards the end of this article) do not change significantly across the set of all known glassformers. Additional details are found in Refs.~[\onlinecite{bib:DSH,bib:DEH2}]. The theory of Ref.~[\onlinecite{bib:DEH1}] first predicted Eq. (\ref{scaling}) with 
\begin{eqnarray}
F(x)= \frac{1}{{\rm erfc}(x)}.
\label{fz}
\end{eqnarray} 
This functional form is consistent with the data \cite{bib:DSH,bib:DEH2}.  
Regardless of theoretical prejudice, the raw experimental collapse of Fig. \ref{fig:collapse} illustrates that, in the broad range of measured $x \equiv \frac{T_{melt}-T}{B T}$ values, Eq. (\ref{scaling}) holds very well (with a function $F$ that is either identical or very close to Eq. (\ref{fz})). Similar to equilibrium transitions, \cite{Guggenheim} the data collapse that we find hints at an underlying universal description of glass formers. This data collapse suggests that, contrary to a long held belief, the equilibrium melting temperature may play a central role in supercooled liquids and glasses. 

\section{Spread in Energy Densities Caused by Supercooling: Intuitive Arguments Rationalizing the Universal Viscosity Collapse}
\label{ref:intuition}

In this Section and those that follow, a function that we will focus on is the probability distribution for the energy density of a supercooled liquid (or glass) at temperature $T$. This function is defined as
\begin{eqnarray}
\label{ptt}
P_{T}(\epsilon') = Tr[\delta(\frac{H}{V}- \epsilon') \rho],
\end{eqnarray}
with $\delta(z)$ a Dirac delta function, $H$ the exact system Hamiltonian (to be explicitly reviewed shortly (Eq. (\ref{atomic})), and $\rho$ the system density matrix just after supercooling.  The central feature that we will wish to motivate is that, unlike regular equilibrium systems, in glasses and supercooled liquids, the distribution $P_{T}(\epsilon')$ may obtain a finite standard deviation $\sigma_{\epsilon}$ in the thermodynamic limit. We will first turn to an intuitive discussion. 

Liquids that are cooled slowly (so that they remain close to equilibrium) may start to crystallize (or veer towards their more general equilibrium solid phase) below a threshold (freezing) transition temperature. By contrast, supercooled liquids have inadequate time for their constituents to crystalize. Thus, a natural viewpoint is that during supercooling multiple low energy structures may start to appear and compete with one another, e.g., \cite{Perim,Confusion} (See Figure \ref{fig:perim}). Indeed, experiments analyzed via Reverse Monte Carlo techniques \cite{RMC}, attest to {\it a distribution} of low energy structures, see, e.g., \cite{st_signartures} in various metrics (including so-called Honeycutt-Andersen indices \cite{HA}). Taken literally, such a broad distribution not only characterizes the frequency of states of low energies but also those in other tail end- states of elevated energy densities. In a somewhat related vein, supercooled liquids indeed exhibit a broad spectrum of spatially non uniform dynamics \cite{DH1,DH2,DH3,DH4,DH5}- some regions of the system are of elevated energy densities by comparison to others. This characteristics is often termed ``dynamical heterogeneity''- meaning that the dynamics are, indeed, not spatially uniform. This spatial distribution qualitatively reflects a related aspect of this widening. This spatial distribution qualitatively reflects a related aspect of this widening. 

Whenever they appear, states (or modes) of high enough energy densities may support long time flow. That is, just as in equilibrium systems, states of energy densities above those of the system at the onset of its freezing point describe the fluid (in which hydrodynamic 
motion occurs). Conversely, states having energy densities that are depressed relative to those of the equilibrium system at its melting point, cannot display flow. These melting (or freezing) cutoffs for hydrodynamic flow are associated with the average of the long time flow velocity over {\it all states} having a well defined energy density. 
Since the equilibrium average is evaluated over all possible configurations of a given energy density, it may be hardly surprising that precisely the same freezing/melting energy density cutoffs will raise their head for the states of the supercooled liquid. In Section \ref{sketch}, we will sketch a rather simple calculation for the long time average velocity in the fluid and its reciprocal, the viscosity, assuming a Gaussian spread of the energy density (a spread that is linear in temperature as governed by the dimensionless factor $B$), and a melting (or freezing) energy density cutoff for having a non-vanishing long time velocity readily yields Eqs. (\ref{scaling},\ref{fz}). 

When examined from an information theoretic perspective, this may not seem surprising. The long range order that persists in a crystal is associated with a low entropy state. It requires little information to describe a crystal. By contrast, the lack of long range correlations in the equilibrium liquid state implies a considerably higher entropy; much more information is required to elucidate the exact state of the system at any time. The supercooled liquid is observed to fall somewhere in a continuum between these two phases. It is natural to conjecture that the information necessary to describe the state of the supercooled liquid (straddling the line between these two limiting equilibrium states) would require data about the underlying equilibrium crystalline and liquid states. A conditioning of the entropy is natural. On broad grounds, maximizing the Shannon entropy $(- \int d \epsilon' P_{T}(\epsilon') \log P_{T}(\epsilon'))$ subject to the constraints of a specified finite standard deviation $\sigma_{\epsilon}$ that enables the system to straddle different equilibrium states and a given average energy density $\epsilon$ leads to a Gaussian distribution $P_{T}(\epsilon')$. In the absence of such a condition of a finite $\sigma_{\epsilon}$, i.e., in the standard equilibrium ({\sf{equil.}}) setting of the canonical ensemble, a maximization of the Shannon entropy distribution leads to a Gaussian distribution for the energy density with a standard deviation $\sigma^{\sf{equil.}}_{\epsilon} = \frac{\sqrt{k_{B} T^{2} C_{v}(T)}}{V} ={\cal{O}}(N^{-1/2})$. Here, $k_{B}$ is the Boltzmann constant, $C_{v} = {\cal{O}}(N)$ is the heat capacity. That is, in equilibrium, the distribution $P_{T}(\epsilon')$ of the energy energy is a Gaussian of a vanishing width in the thermodynamic ($N \to \infty$) limit. In Section \ref{sketch}, we will invoke precisely such a normal distribution that, as emphasized above, maximizes the Shannon entropy with a constrained standard deviation 
$\sigma_{\epsilon}$ that (unlike $\sigma^{\sf{equil.}}_{\epsilon}$) is of a {\it finite value}.

\section{Rigorous bounds on the energy distribution widths in systems with a varying temperature}
\label{bound}
At first glance, the results that we next review might seem to be abstract mathematical statements that are conceptually decoupled and have little to do with the very vividly simple considerations of Section \ref{ref:intuition} and of Figure \ref{fig:perim}. These considerations are indeed far afield from simple classical pictures. However, we feel that they provide a glimpse into the generality of the features motivated in Section \ref{ref:intuition} in very diverse settings. Following Ref.[\onlinecite{QZN}], we will now demonstrate that whenever a system is cooled or heated in such a way that its energy density (or measured temperature) varies at a finite rate, then the energy density (even when this density is globally evaluated over the entire macroscopic system) must exhibit a finite width. Naively, such a general statement concerning global averages in a macroscopic systems may appear impossible in systems with local interactions. Indeed, by the central limit theorem, the energy density $\epsilon$ (or any other intensive quantity) can only exhibit fluctuations that diminish as ${\cal{O}}(N^{-1/2})$. In particular, in the thermodynamic limit, the standard deviation associated with $\epsilon$ may be expected to vanish. Such arguments appealing to the central limit theorem implicitly assume that the system is in equilibrium or that it displays no entanglement. 

Although it is not often appreciated, typical thermal states are, in fact, highly entangled. In their ground states ($T=0$), most systems display ``area law'' entanglement entropies. At all other positive temperatures, general systems may exhibit volume law entanglement \cite{volume1,volume2,volume3,volume4}.  A trivial yet important point is that although equilibrium states exhibit (by their defining nature) sharp intensive state variables (e.g., energy or number densities), measurements of other generic quantities need not always yield sharp outcomes with a vanishing standard deviation. 

As a pedagogical illustration \cite{QZN}, consider a situation in which a cooled or heated subsystem (governed, in the absence of external cooling or heating, by
a Hamiltonian $H$) is part of a larger closed system (defined by the Hamiltonian $\tilde{H}$). When the subsystem is cooled or heated such that the rate of change of its measured temperature $dT/dt$  (or energy density) is finite, the derivative of energy must be extensive, $\frac{dE}{dt} = \langle \frac{d H_{H}}{dt} \rangle = {\cal{O}}(N)$. Here, $H_{H} \equiv e^{i\tilde{H} t/\hbar} H e^{-i \tilde{H} t /\hbar}$ is the the Hamiltonian of the subsystem in the Heisenberg representation
and the average $\langle  - \rangle$ is evaluated with the probability density matrix of the initial $(t=0)$ state (which we may consider to be an equilibrium state). 
By Heisenberg's equation, $ \langle \frac{d H_{H}}{dt} \rangle = \frac{i}{\hbar} \langle [\tilde{H}, H_{H}] \rangle$. We next invoke the generalized uncertainty inequalities to obtain $|\langle [\tilde{H}, H_{H}] \rangle| \le 2 \sigma_{\tilde{H}} \sigma_{H_{H}}$. Here, $\sigma_{H_{H}}$ and $\sigma_{\tilde{H}}$ respectively denote the widths (or ``uncertainties'') in the open subsystem and closed global system Hamiltonians. 
Putting these pieces together,
\begin{eqnarray}
\label{HH}
 \sigma_{\tilde{H}} \sigma_{H_{H}} \ge \frac{\hbar}{2} \Big|\frac{dE}{dt} \Big|.
\end{eqnarray} 
Now if, regardless of the special initial state that we prepare associated with the specific cooling/heating protocol of the subsystem, the closed system exhibits ergodicity at long times such that long time averages become equilibrium averages with $\tilde{H}$ then the uncertainty $\sigma_{\tilde{H}} = {\cal{O}}(1)$ is system size independent. That is, closed systems at equilibrium may be analyzed via the micro-canonical ensemble (in which $\sigma_{\tilde{H}} = {\cal{O}}(1)$). Since the uncertainty in the energy in a closed system is time independent, for systems that equilibrate at long times, $\sigma_{\tilde{H}} = {\cal{O}}(1)$ at all times. From Eq. (\ref{HH}), we then have that the uncertainty in the energy density \cite{QZN}, 
\begin{eqnarray}
\label{eq:bound}
\sigma_{\epsilon} =  \frac{\sigma_{H_{H}}}{V}= {\cal{O}}(1).
\end{eqnarray}
Thus, having a finite width of the energy density is, essentially, inescapable. This general conclusion may be made vivid by exact calculations on various model systems such as generic Heisenberg ferromagnets or antiferromagnets on arbitrary lattices that are quenched by altering external magnetic fields \cite{QZN}. The bounds of Eqs. (\ref{HH}, \ref{eq:bound}) draw attention to pervasive broad energy density distributions in systems that are forcefully driven away from equilibrium by external cooling or heating. If the initial ($t=0$) state is an equilibrium state then the uncertainty of the energy density in that initial state, $\frac{\sigma_{H}}{V} = {\cal{O}}(N^{-1/2}) \to_{N \to \infty} 0$.
However, relative to that same initial state, the operator $H_{H}(t)$ has a large variance at times $t$ during which the system is cooled or heated.   
For systems with local Hamiltonians expressible as $H_{H}(t) = \sum_{R} {\cal{H}}_{R}(t)$ with 
${\cal{H}}_{R}(t)$ denoting local Hamiltonians associated with a spatial location $R$ (examples of which can
be readily written \cite{QZN}), the standard deviation of the energy density reads
\begin{eqnarray}
\sigma_{\epsilon}^{2} = \frac{1}{N^{2}} \sum_{R,R'} (\langle {\cal{H}}_{R} {\cal{H}}_{R'} \rangle - \langle {\cal{H}}_{R} \rangle \langle {\cal{H}}_{R'} \rangle) \equiv \frac{1}{N^{2}} \sum_{R,R'} G(R,R').
\end{eqnarray}
A finite width $\sigma_{\epsilon}$ of the energy density then implies that even for $R$ and $R'$ arbitrarily far apart, the average connected correlation function $G(R,R')$ does not vanish. Thus, cooling or heating may trigger measurable long range correlations in an initial, seemingly trivial, thermal state. Once cooling or heating ceases (so that $\frac{d \epsilon}{dt} =0$), the lower bound on $\sigma_{\epsilon}$ vanishes. If there are no impediments for the system to return to equilibrium then it may have a vanishing standard deviation of the energy density shortly after cooling/heating stops. However, in systems that ``get stuck'' and cannot fully equilibrate, a finite standard deviation of the energy density may persist to long times. Strong hints that this is indeed the case are afforded by the deviations of the average values of various observables in supercooled liquids and glasses vis a vis their average values in equilibrium solids and liquids. By its defining nature, the expectation value of any observable ${\cal Q}$ in the microcanonical  ({\sf m.c.}) ensemble is the {\em average} value of ${\cal Q}$ over the ${\cal{N}}(E,\Delta)$ eigenstates that lie in the energy window $E \le E_{n} \le E + \Delta$, i.e, 
\begin{eqnarray}
\label{QQm}
\langle {\cal Q} \rangle_{\sf m.c.} \equiv \frac{1}{{\cal{N}} (E,\Delta)} \sum_{E\le E_{n} \le E + \Delta} \langle \phi_n | {\cal Q} | \phi_n \rangle.
\end{eqnarray} 
The width of the energies over which this sum extends $\Delta = {\cal{O}}(1)$ is system size independent. In the thermodynamic limit, $\Delta/V \to 0$ and the distribution of energy densities appearing in the sum of Eq. (\ref{QQm}) becomes a delta-function. Similarly, as we discussed earlier, also for open equilibrated systems (e.g., the canonical ensemble which we most commonly allude to), the spectral width of the probability density matrix $\sigma^{\sf{equil.}} ={\cal{O}}(N^{-1/2})$ also tends to zero as $N \to \infty$ (as it must for the shape state variable $\epsilon$). 

 Now we turn to experimental facts. The disparity between the values of measured observables in supercooled liquids or glasses vis a vis their values (given by Eq. (\ref{QQm})) in truly equilibrated solids implies that $P_{T}(\epsilon')$ cannot be a distribution of vanishing width 
$\sigma_{\epsilon}$ in the $N \to \infty$ limit. This is so since if $\sigma_{\epsilon}$ vanished, the average values of all observables in supercooled liquids and glasses would be equal to those in truly equilibrated solids and liquids. Thus, for supercooled liquids and glasses, the distribution $P_{T}(\epsilon')$ of energy densities must, instead, acquire a finite width $\sigma_{\epsilon}$. 

Indeed, excusing ``many body localized'' states \cite{MBL1,MBL2,MBL3,MBL4,MBL5,MBL6,noMBL}, the expectation values of general observables in typical states of a given energy density may be identical to those found in equilibrium. That is, the equilibrium averages of general physical observables at a certain temperature or energy density are equal to expectation values of the same observables in typical states of the same prescribed energy density. Non-equilibrium behaviors can appear only if, somehow, these individual states (or special sets of such states) are extremely special and display expectation values different from the ensemble averages over all states of a given energy density. Thus, the appearance  of a finite energy density width constitutes a very general mechanism for exhibiting non-equilibrium behaviors. The central hypothesis of our approach is that this simple broadening is the sole principal feature separating supercooled liquids and glasses from their equilibrium counterparts. Combined with Eq. (\ref{eq:bound}), the intuitive considerations of Section \ref{ref:intuition}, elicit us to contemplate this simple possibility. 
For completeness, we further mention that broad distributions $\sigma$ may also appear in ``non self averaging''  disordered classical systems \cite{derida,amnon,eitan,per} and, rather trivially, in classical systems with interactions that are of unbounded spatial range or strength. 

\section{Sketch of the long time averages of general observables and the viscosity}
\label{sketch}
Armed with the proof of principle rigorous bounds of Section \ref{bound}, we now return to the pictorial ideas outlined in Section \ref{ref:intuition} and couch these in a broader context. The basic strategy that we will follow concerns the logical outcome of a rather tautological statement concerning completeness. The same {\it complete set} of many body atomic states (whether quantum mechanical eigenstates or classical microstates) describes both (1) the supercooled liquids and glasses and (2) the equilibrium systems. The completeness of these states implies that all observable properties of glasses may be related (via a linear transformation) to the values of the very same observables when measured in equilibrium systems. Since the equilibrium systems undergo a transition only at their melting temperature, the equilibrium melting temperature also plays a vital role in supercooled liquids. More precisely, the eigenstates of the same basic many body Hamiltonian,
 \begin{eqnarray}
\label{atomic}
H= -\sum_{i} \frac{\hbar^{2}}{2M_{i}} \nabla^{2}_{R_{i}} - \sum_{j} \frac{\hbar^{2}}{2m_{e}} \nabla_{r_{j}}^{2} - \sum_{i,j} \frac{Z_{i}e^{2}}{|R_{i}-r_{j}|} \nonumber
\\  + \frac{1}{2} \sum_{i \neq i'} \frac{Z_{i} Z_{i'} e^{2}}{|R_{i} - R_{i'}|} + \frac{1}{2} \sum_{j\neq j'} \frac{e^{2}}{|r_{j} - r_{j'}|}.
\end{eqnarray}
governing both systems of types (1) and (2) exhibit a ``phase transition'' and are non-analytic at the energy densities associated with melting. 
The Hamiltonian $H$ describes these systems on all scales of experimental interest. In Eq. (\ref{atomic}), $M_{i}, R_{i},$ and $Z_{i}e$ are, correspondingly, the mass, position, and charge of the $i-$th nucleus, 
while $r_{j}$ is the location of the $j-$th electron (whose mass and charge are $m_{e}$ and $(-e)$ respectively). The spectral problem posed by $H$ is, of course, nontrivial. We recall however two points. (i) {\em Empirically}, the equilibrium averages of various observables display singularities only at phase transitions. (ii) These equilibrium expectation values (in, e.g., the microcanonical or canonical ensembles) are averages over all eigenstates having the same energy density. Taken together (i) and (ii) imply that typical eigenstates of Eq. (\ref{atomic}) display non-analyticities at the very same energy densities at which phase transitions occur experimentally. Making contact with the discussion in the Introduction, at the energy densities at which equilibrium transitions appear, topological melting occurs.
The typical eigenstates at energies above and below the melting transition are of an inherently different character and reflect the characteristics of the equilibrium 
liquid and the solid. We reiterate that the only the energy densities of pertinence to non-analyticities of the eigenstates of $H$ are those associated with experimental phase transitions. This, then, suggests that the viscosity and relaxation times of all glass formers should collapse onto a universal curve with the only important temperature scale indeed being that of equilibrium melting. We wish to explicitly further expand on outcomes of the logical steps that we have just invoked. 
Towards this end, we reiterate that the microcanonical ensemble average of Eq. (\ref{QQm}) is (for standard systems that equilibrate and obey ensemble equivalence) equal to the measured value of ${\cal Q}$ in thermal equilibrium $\langle {\cal Q}(\epsilon) \rangle_{eq}$ at the associated energy density $\epsilon$. Given this $\langle {\cal Q}(\epsilon) \rangle_{\sf m.c.} = \langle {\cal Q}(\epsilon) \rangle_{eq}$ equivalence, we see that an {\em average} of any observable over eigenstates $\{| \phi_{n} \rangle\}$ of energy density $\epsilon$ assumes its value in a typical \newline
{\sf (a)} state of an equilibrium solid phase of energy density $\epsilon$ when $\epsilon < \epsilon_{melt}$, or of an \newline
{\bf (b)} equilibrium state of the liquid (or higher energy density phases) when $\epsilon > \epsilon_{melt}$. 
This conclusion regarding the average (or typical) eigenstates of general energy densities $\epsilon$ reaffirms the intuitive statements made in Section \ref{ref:intuition}. We {\em do not need} to address the tall order of diagonalizing Eq. (\ref{atomic}) is order to ascertain the exact average of  $\langle \phi_n | {\cal Q} | \phi_n \rangle$ over the eigenstates $| \phi_{n} \rangle$ of $H$ in a narrow energy window $[E,E+ \Delta]$. Our basic premise is that the {\it experimental results on the equilibrium system defined by $H$ essentially solve the spectral problem for us}. If particular, since when subject to an infinitesimal external stress there is no long time flow in an {\it equilibrium} solid, we explicitly deduce a particular corollary of {\sf{(a)}}. Namely, \newline
{\sf (c)} eigenstates with an energy density $\epsilon < \epsilon_{melt}$ (more precisely, eigenstates of energy density lower than the equilibrium ``liquidus'' value below which solid inclusions appear) do not, on average, support long time hydrodynamic flow for infinitesimal external stress.\newline
A straightforward calculation \cite{bib:DEH1} illustrates that the long time average ${\cal{Q}}_{\infty}$ of any quantity or operator (${\cal Q}$ whose experimental equilibrium value $\langle {\cal Q}(\epsilon) \rangle_{eq}$ is largely temperature (or energy density $\epsilon'$) dependent is given by 
\begin{eqnarray}
\label{QQT}
{\cal{Q}}_{\infty} \equiv \lim_{\tilde{{\cal{T}}} \to \infty} \frac{1}{\tilde{\cal{T}}} \int_{t_{final}}^{t_{final} + {\tilde{\cal{T}}}} dt' ~Tr[\rho(t'){\cal{Q}}] = \int d \epsilon' P_{T}(\epsilon')
\langle {\cal Q}(\epsilon') \rangle_{eq}.
\end{eqnarray}
In Eq. (\ref{QQT}), $t_{final}$ is the time at which the influence of external quenching halts and the supercooled liquid evolves as a closed system with the Hamiltonian of Eq. (\ref{atomic}). Eq. (\ref{QQT})
 is derived by a simple sequence of steps. We write the Schrodinger picture density matrix $\rho(t) = e^{-iH(t-t_{final})/\hbar} \rho e^{iH(t-t_{final})/\hbar}$, insert a resolution of the identity with the eigenstates of $H$
 (or, more generally, resolving the identity via the common eigenstates of $H$ and ${\cal{Q}}$ when the eigenstates of $H$ are degenerate), replace the latter sum over all eigenstates of $H$ (or of common eigenstates  of both $H$ and $Q$) by an integral over $\epsilon'$ (with any additional sum over eigenstates of ${\cal{Q}}$ if degeneracy exists) by using Eq. (\ref{QQm}), invoke the equivalence 
 $\langle {\cal Q}(\epsilon) \rangle_{\sf m.c.} = \langle {\cal Q}(\epsilon) \rangle_{eq}$, note that the ratio $(\int_{0}^{\tilde{\cal{T}}} dt'~e^{i \omega_{nm} t'})/{\tilde{{\cal{T}}}}$ tends to zero for large ${\tilde{{\cal{T}}}}$ whenever the difference in energies amongst eigenstates appearing in the above resolution of the identity is non-vanishing (i.e., the above ratio tends to zero when $\omega_{nm} \equiv (E_{n}-E_{m})/\hbar \neq 0$), and lastly insert Eq. (\ref{ptt}). We now use our conclusion {\sf (c)} and Eq. (\ref{QQT}) to predict the measured viscosity of supercooled liquids. One of the oldest methods of determining the viscosity is to measure the terminal velocity of a sphere dropped into the fluid. By Stokes' law, the terminal velocity of a sphere of radius $R$ and mass density $\rho_{sphere}$ in a fluid of viscosity $\eta$ and mass density $\rho_{fluid}$ is
\begin{eqnarray}
\label{stokes}
v_{\infty} =  \frac{2}{9} \frac{\rho_{sphere} - \rho_{fluid}}{\eta} g R^{2}.
\end{eqnarray}
On the other hand, if the state of the supercooled system at a temperature $T$ emulates a distributed average (with a probability distribution $P_{T}(\epsilon')$) of the equilibrium result 
(or typical eigenstate) for an energy density $\epsilon'$ then, by invoking Eq. (\ref{QQT}), 
\begin{eqnarray}
v_{\infty}  = \int d \epsilon' ~P_{T}(\epsilon')~ v^{eq}_{\infty}(\epsilon').
\label{vev}
\end{eqnarray}
Here, $v^{eq}_{\infty}(\epsilon')$ is the terminal velocity of the sphere when the same system is at equilibrium with an energy density $\epsilon'$. 
Since the terminal velocity of the sphere will vanish in the solid phase (when $\epsilon' < \epsilon_{melt}$), the integral of Eq. (\ref{vev}) may be performed
from $\epsilon'=\epsilon_{melt}$ to $\epsilon'=\infty$). If the distribution $P_{T}(\epsilon')$ has most of its weight at energies below $\epsilon_{melt}$
and the equilibrium terminal velocity $v^{eq}_{\infty}(\epsilon')$ is a weak function of temperature of energy densities above melting (i.e., $v^{eq}_{\infty}(\epsilon' >\epsilon_{melt}) \sim v^{eq}_{\infty}(\epsilon^{+}_{melt})$) then
\begin{eqnarray}
v_{\infty}  = v^{eq}_{\infty}(\epsilon^{+}_{melt}) \int _{\epsilon_{melt}}^{\infty} d \epsilon'~P_{T}(\epsilon').
\label{vev'}
\end{eqnarray}
We consider a Gaussian $P_{T}(\epsilon')$ (a distribution that maximizes the Shannon entropy given a finite standard deviation $\sigma_{\epsilon}$),
of standard deviation $\sigma_{\epsilon}$ such that the ratio  
\begin{eqnarray}
\label{eq:B}
B \equiv \frac{\sigma_{\epsilon} (T_{melt}-T) \sqrt{2}}{T(\epsilon_{melt}- \epsilon)}
\end{eqnarray}
is, approximately, constant in the temperature regime ($T_{g} \lesssim T<T_{melt}$) of experimental relevance. Combining Eq. (\ref{stokes}) (applied to both the viscosity of the supercooled liquid and the viscosity of the equilibrium system
at energy density $\epsilon'$) with Eqs. (\ref{vev'}, \ref{eq:B}) produces Eqs. (\ref{scaling}, \ref{fz}) \cite{bib:DEH1}. 
Although, in the rather trivial derivation above, we have invoked the eigenstate decomposition of general states, one may provide different verbal rationalizations to the simple relations of Eqs. (\ref{vev'},\ref{eq:B}) that led us to suggest and find the collapse of Fig. (\ref{fig:collapse}). The key point is that a Gaussian distribution of modes that contribute to long time flow will rather robustly yield Eqs. (\ref{scaling}, \ref{fz}) that is indeed rather universally obeyed as we show in Fig. \ref{fig:collapse} and discussed in depth in \cite{bib:DSH,bib:DEH2}. The above calculation may be repeated, {\it mutatis mutandis}, for other relaxation times (such
as those associated with dielectric response relaxations) and a multitude of thermodynamic and structural properties. Elsewhere, we will show
that dielectric relaxation data collapse on the very same universal curve shown in Fig. \ref{fig:collapse}. 
In fact, the very same unique probability distribution $P_{T}$ should universally relate numerous measurable properties of
the glass to those of equilibrium systems at various energy densities $\epsilon'$. 

To close the circle of ideas that we started to explore at the beginning of this article, we remark that superposing different crystalline (or other) structures of varying energy densities may generate non-uniform amorphous structures (see, e.g., Fig.  (\ref{fig:perim})). This is the structural counterpart of decomposing ill-organized configurations into ordered (``crystalline like'') periodic Fourier modes. 

\section{Conclusions}
In summary, we outlined a new approach to the glass transition. Our central proposal is that when the {\it usual topological defect driven transition} from the equilibrium solid to the liquid is merely smeared by a normal probability distribution of a finite width, then the resulting theory will reproduce quite well the phenomenology of glasses and supercooled liquids. That is, the sole difference between our approach and the usual equilibrium transition is that the probability distribution of the energy density governing supercooled liquids and glasses is a normal distribution of a finite width (instead of a Gaussian of width scaling as ${\cal{O}}(N^{-1/2})$ in finite temperature 
{\it equilibrium} systems). A prediction of our theory is that the viscosity and dielectric relaxation times collapse on a simple curve (given by Eqs. (\ref{scaling}, \ref{fz})). This conjecture is indeed satisfied for all glass formers (as seen in Fig. (\ref{fig:collapse})). In a similar vein, numerous other experimental observables should be governed by this same distribution function. We underscore that the empirical collapse of Fig. (\ref{fig:collapse}) holds universally regardless of the theoretical considerations that led us to it.  

\section{Acknowledgments}
 ZN and NBW gratefully acknowledge support from the National Science Foundation under grant number NSF 1411229. FSN thanks the German Collaborative Research Center SFB 1143. We further wish to thank the organizers of the workshop on ``Topological Phase Transitions and New Developments''.

\appendix{Numerical fit values} 

In the table that follows, the parameter $B$ in the collapse of  Eq. \ref{scaling} and Fig. \ref{fig:collapse} is provided. 
Further detail concerning the analysis leading to these values and many further aspects (both empirical and theoretical) appear in \cite{,bib:DSH,bib:DEH2}. 
  
\begin{table*}[t]
\centering
\caption{Values of Relevant Parameters for all liquids studied}
\begin{tabular}{*{4}{@{\hskip 0.4in}c@{\hskip 0.4in}}} 
\emph {Composition} \quad & \emph{} $B$ & \emph{$T_{melt}$ [K]} &\emph{$\eta(T_{melt})$ [Pa*s]} \\
\colrule
BS2 & 0.157129 & 1699 & 5.570596   \\
Diopside & 0.134328 & 1664 & 1.5068     \\
LS2 & 0.170384 & 1307 & 22.198     \\
OTP & 0.069685 & 329.35 & 0.02954    \\
Salol & 0.087192  & 315 & 0.008884   \\
Anorthite & 0.131345 & 1823 & 39.81072   \\
$Zr_{57}Ni_{43}$ & 0.234171 & 1450 & 0.01564    \\
$Pd_{40}Ni_{40}P_{20}$ & 0.154701 & 1030 & 0.030197   \\
$Zr_{74}Rh_{26}$ & 0.187851 & 1350 & 0.03643    \\
$Pd_{77.5}Cu_6Si_{16.5}$ &  0.124879 & 1058 & 0.0446     \\
Albite & 0.103344 & 1393 & 24154952.8 \\
$Cu_{64}Zr_{36}$ &  0.142960 & 1230 & 0.021      \\
$Ni_{34}Zr_{66}$ & 0.209359 & 1283 & 0.0269     \\
$Zr_{50}Cu_{48}Al_{2}$ &0.167270 & 1220 & 0.0233     \\
$Ni_{62}Nb_{38}$ & 0.109488  & 1483 & 0.042      \\
Vit106a &0.133724  & 1125 & 0.131      \\
$Cu_{55}Zr_{45}$ & 0.144521 & 1193 & 0.0266     \\
$H_2O$ & 0.133069 & 273.15 & 0.001794   \\
Glucose & 0.079455 & 419 & 0.53       \\
Glycerol &  0.108834 & 290.9 & 1.9953     \\
$Ti_{40}Zr_{10}Cu_{30}Pd_{20}$ &0.185389  & 1279.226 & 0.01652    \\
$Zr_{70}Pd_{30}$ &  0.21073 & 1350.789 & 0.02288    \\
$Zr_{80}Pt_{20}$ & 0.169362 & 1363.789 & 0.04805    \\
NS2 & 0.134626 & 1147 & 992.274716 \\
$Cu_{60}Zr_{20}Ti_{20}$ & 0.103380 & 1125.409 & 0.04516    \\
$Cu_{69}Zr_{31}$         & 0.157480  & 1313     & 0.01155    \\
$Cu_{46}Zr_{54}$         &  0.156955 & 1198     & 0.02044535 \\
$Ni_{24}Zr_{76}$         & 0.244979 & 1233     & 0.02625234 \\
$Cu_{50}Zr_{42.5}Ti_{7.5}$  & 0.148249 & 1152     & 0.0268     \\
D Fructose       &  0.050124 & 418      & 7.31553376 \\
TNB1             &  0.07567 & 472      & 0.03999447 \\
Selenium         & 0.130819  & 494      & 2.9512     \\
CN60.40          & 0.149085  & 1170     & 186.2087   \\
CN60.20          & 0.161171  & 1450     & 12.5887052 \\
$Pd_{82}Si_{18}$         &  0.137623 & 1071     & 0.03615283 \\
$Cu_{50}Zr_{45}Al_{5}$     & 0.118631   & 1173     & 0.03797    \\
$Ti_{40}Zr_{10}Cu_{36}Pd_{14}$ &  0.137753 & 1185     & 0.0256     \\
$Cu_{50}Zr_{50}$       & 0.166699 & 1226     & 0.02162    \\
Isopropylbenzene &   0.073845 & 177      & 0.086      \\
ButylBenzene     & 0.085066  & 185      & 0.0992     \\
$Cu_{58}Zr_{42}$         &  0.131969 & 1199     & 0.02526    \\
Vit 1 & 0.111185 & 937 & 36.59823 \\
Trehalose & 0.071056  & 473 & 2.71828 \\
Sec-Butylbenzene & 0.080088 & 190.3 & 0.071 \\
$SiO_2$ & 0.090948 & 1873 & 1.196x$10^{8}$ \\
\hline       
\end{tabular}
\label{table}
\end{table*}



\end{document}